\documentclass[letterpaper,usenatbib,usegraphicx]{mn2e}
\usepackage[dvipdfm]{hyperref}
\usepackage{amssymb}
\usepackage{amsfonts}

\newcommand {\ie} {{\it i.e.}}

\newcommand {\eg} {{\it e.g.}}

\newcommand {\be} {\begin{equation}}
\newcommand {\ee} {\end{equation}}
\newcommand {\bea} {\begin{eqnarray}}
\newcommand {\eea} {\end{eqnarray}}

\defcitealias{2009MNRAS.392.1205N}{Paper I}

\title[The Efficiency of Reconfinement Shocks]{Dissipation Efficiency of Reconfinement Shocks in Relativistic Jets}
\author[Nalewajko]{Krzysztof Nalewajko\thanks{E-mail: knalew@jila.colorado.edu}\\
JILA, University of Colorado and National Institute of Standards and Technology, 440 UCB, Boulder, CO 80309, USA\\
Nicolaus Copernicus Astronomical Centre, Bartycka 18, 00-716 Warsaw, Poland}

\begin{document}

\maketitle

\begin{abstract}
We calculate the dissipation efficiency of relativistic reconfinement shocks. Building on previous work (Nalewajko \& Sikora 2009), we consider different distributions of the external pressure. The average dissipation efficiency $\epsilon_{\rm diss}$ is a function of the product of two parameters -- the jet Lorentz factor $\Gamma_{\rm j}$ and the reconfinement angle $\Theta_{\rm r}$, which is related to the opening angle $\Theta_{\rm j}$ and the external pressure index $\eta$. The spatial distribution of the dissipation rate strongly depends on $\eta$. We discuss the significance of these results for the properties of relativistic jets in gamma-ray bursts and active galactic nuclei and propose that reconfinement shocks may explain a very high dissipation efficiency of the former and a moderate dissipation efficiency of the latter. Finally, we estimate the dissipation efficiency of the reconfinement shock associated with the quasi-stationary knot HST-1 in the jet of radio galaxy M87 and show that it is roughly consistent with the observational constraints.
\end{abstract}


\section{Introduction}
\label{sec_intro}


Relativistic jets stand behind the brightest cosmic phenomena: gamma-ray bursts (GRBs) and blazars, a subclass of active galactic nuclei (AGNs). Their extreme isotropic luminosities, up to $\sim 10^{53}\;{\rm erg\;s^{-1}}$ for the former \citep[GRB~080319B;][]{2008Natur.455..183R} and up to $\sim 10^{50}\;{\rm erg\;s^{-1}}$ for the latter \citep[3C~454.3;][]{2011ApJ...733L..26A}, cannot be plausibly explained without the relativistic Doppler effect. However, even taking this into account, in order for this radiation to be produced in the co-moving reference frame, a substantial fraction of the jet mechanical power, including the particle rest energy flux, needs to be dissipated, then transferred into a population of ultra-relativistic particles in a non-thermal acceleration process and finally radiated away through non-thermal radiative mechanisms. The total efficiency of these processes can be estimated observationally if the total jet power is known. In the case of GRBs, it can be well constrained by energetics of the afterglow phase and the radiative efficiency of the prompt phase has been claimed reach values up to $\sim 90\%$ \citep{2007ApJ...655..989Z}. In the case of blazars, these estimates are less certain, but typical values for their luminous class of Flat Spectrum Radio Quasars (FSRQs) are $\sim 10\%$ \citep{2008MNRAS.385..283C}. Regardless of efficiencies of the particle acceleration and radiative processes, efficiency of the energy dissipation in relativistic jets must be at least comparable to these observational constraints.


The most widely discussed means of energy dissipation in relativistic jets are shock waves, magnetic reconnection and instabilities.
Shocks can arise within a jet when two regions propagating with substantially different bulk velocities collide with each other. Such internal shocks provided the basic framework for theoretical models of blazars \citep[\eg,][]{1979ApJ...232...34B} and GRBs \citep[\eg,][]{1994ApJ...430L..93R}.
However, these models have been questioned on the grounds that they cannot account for required dissipation efficiencies. This is especially clear in the case of GRBs, for which several alternative models have been recently proposed, based on magnetic reconnection \citep[\eg,][]{2010arXiv1011.1904M,2011ApJ...726...90Z} or relativistic turbulence \citep[\eg,][]{2009MNRAS.394L.117N}. In the case of blazars, detailed calculations showed that a substantial contrast of initial Lorentz factors must be assumed \citep{2001MNRAS.325.1559S}. However, the occurrence of such a velocity contrast cannot be verified with current models of jet formation and acceleration. Moreover, if the velocity modulations are related to processes at the black hole horizon scale, the internal shocks model predicts a particular length scale, a fraction of a parsec, over which such shocks develop. There are now several arguments for the bulk emission of luminous blazars being produced at much larger distances from the central black hole \citep[\eg,][]{2008ApJ...675...71S,2011ApJ...726L..13A}.


The other possibility for the shocks is that they result from the interaction between the jet and its environment. In the case of GRBs, the jet is a transient phenomenon and has to plough through its host star and the interstellar medium, forming an external shock that dominates during the afterglow phase \citep[\eg,][]{1997ApJ...476..232M}. In the case of blazars, the jet is relatively persistent and propagates roughly along a tunnel drilled over time, so perpendicular external shocks are not usually considered. However, the external medium can exert a substantial pressure on the jet boundary, forcing it to recollimate and triggering a reconfinement shock.


Reconfinement shocks were first discussed by \cite{1983ApJ...266...73S} in the context of the kpc-scale jet of the radio galaxy NGC~315. The first analytical models were introduced by \cite{1989RMxAA..17...65C} in the non-relativistic regime applicable to the jets of young stellar objects (YSOs) and by \cite{1997MNRAS.288..833K} in the relativistic regime. A unique signature of reconfinement shocks is that, unless the jet or external medium parameters vary significantly on the dynamical time scale, they would be observed as stationary patterns. \cite{1988ApJ...334..539D} interpreted a stationary knot in the pc-scale jet of radio quasar 4C~39.25 as a nozzle of the reconfinement shock. More recently, a stationary\footnote{After the 2005 outburst this knot is no longer stationary and propagates with an apparent velocity of $\sim 2.7 c$ \citep{2010arXiv1010.4170G}.} knot HST-1 has been discovered in the jet of radio galaxy M87 at the $100\;{\rm pc}$ scale \citep{1999ApJ...520..621B} and subsequently it underwent a spectacular multiwavelength outburst \citep[\eg][]{2006ApJ...640..211H}. \cite{2006MNRAS.370..981S} showed that the association of this feature with a reconfinement shock is consistent with both the properties of the host galaxy and the estimated jet power. However, short variability time scales required a very compact emitting region. \cite{2009ApJ...699.1274B} showed that efficient focusing of the shocked jet flow is possible, but requires substantial cooling of the post-shock plasma. Reconfinement shocks were also studied in the context of GRBs \citep{2007ApJ...671..678B}.


The problem of dissipation efficiency of relativistic reconfinement shocks was first studied in \cite{2009MNRAS.392.1205N}, hereafter \citetalias{2009MNRAS.392.1205N}. It was found that the dissipation efficiency $\epsilon_{\rm diss}$ depends strongly on the product of the jet Lorentz factor $\Gamma_{\rm j}$ and the opening angle $\Theta_{\rm j}$. Here, we generalise this result, taking into account different distributions of the external pressure. We also show how this result can be applied to both GRBs and the jets of active galactic nuclei. Because GRB jets are characterised by wide opening angles, reconfinement shocks provide a natural explanation of their high radiative efficiency in the prompt phase. In AGN jets, the efficiency of reconfinement shocks is much lower, because collimation by a continuous medium limits the opening angle. We also estimate the efficiency of the reconfinement shock associated with the HST-1 knot in the jet of M87 and show that it is roughly consistent with the observed luminosity of this radio galaxy.


In Section \ref{sec_mod}, we present our simple model of the structure of relativistic reconfinement shocks. The dependence of the dissipation efficiency on model parameters is discussed in Section \ref{sec_res}. In Section \ref{sec_dis}, we discuss the applications of these results to astrophysical relativistic jets. Conclusions are given in Section \ref{sec_con}.

\section{The model}
\label{sec_mod}

\begin{figure}
  \includegraphics[width=\columnwidth]{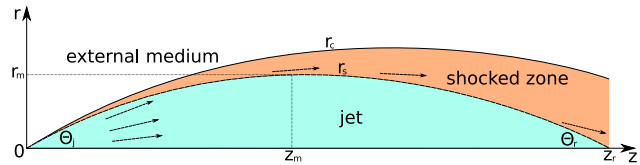}
  \caption{Geometric parameters of our reconfinement shock model. The jet is symmetric around the $z$ axis and propagates from its origin at $z=0$ towards the reconfinement point at $z=z_{\rm r}$. The opening angle $\Theta_{\rm j}$, the reconfinement angle $\Theta_{\rm r}$ and the maximum jet width $r_{\rm m}$ are indicated.}
  \label{fig_scheme}
\end{figure}

Reconfinement shocks result from the interaction between a jet and its surrounding medium. The simplest model of such a problem involves a cold, unmagnetized, spherically symmetric jet of Lorentz factor $\Gamma_{\rm j}$, opening angle $\Theta_{\rm j}$ and total power $L_{\rm j}$; and a static medium of pressure distribution given by $p_{\rm e}(z)\propto z^{-\eta}$, where $\eta<2$ is a constant and $z$ is the coordinate measured along the jet axis. Figure \ref{fig_scheme} shows the geometric parameters of the reconfinement shock front of radius $r_{\rm s}(z)$, including reconfinement length $z_{\rm r}$, reconfinement angle $\Theta_{\rm r}$ and maximum jet width $r_{\rm m}$; as well as the contact discontinuity of radius $r_{\rm c}(z)$. Parameters measured immediately upstream and downstream of the reconfinement shock are denoted with subscripts 'j' and 's', respectively.

The system of shock jump equations is
\bea
\label{eq_mod_shock1a}
\beta_{\rm s}\cos(\theta_{\rm s}-\alpha_{\rm s})&=&\beta_{\rm j}\cos(\theta_{\rm j}-\alpha_{\rm s})\,,\\
\label{eq_mod_shock1b}
u_{\rm s}\rho_{\rm s}\sin(\theta_{\rm s}-\alpha_{\rm s})&=&u_{\rm j}\rho_{\rm j}\sin(\theta_{\rm j}-\alpha_{\rm s})\,,\\
\label{eq_mod_shock1c}
u_{\rm s}^2w_{\rm s}\sin^2(\theta_{\rm s}-\alpha_{\rm s})+p_{\rm s}&=&u_{\rm j}^2\rho_{\rm j}c^2\sin^2(\theta_{\rm j}-\alpha_{\rm s})\,,\\
\label{eq_mod_shock1d}
\Gamma_{\rm s}u_{\rm s}w_{\rm s}\sin(\theta_{\rm s}-\alpha_{\rm s})&=&\Gamma_{\rm j}u_{\rm j}\rho_{\rm j}c^2\sin(\theta_{\rm j}-\alpha_{\rm s})\,,
\eea
where $\beta=v/c$ is the dimensionless velocity, $u=\Gamma\beta$ is the dimensionless four-velocity, $w=\rho c^2+p+e$ is the relativistic enthalpy in the comoving frame, $\rho$ is the mass density, $e$ is the thermal energy density, $\alpha_{\rm s}$ is the inclination of the shock front with respect to the jet axis and $\theta_{\rm j,s}$ are the inclinations of the velocity vectors.
It is assumed that $p_{\rm j}=0$. Given all the parameters of the upstream plasma, this system can be solved when the post-shock pressure $p_{\rm s}$ is given. In \citetalias{2009MNRAS.392.1205N}, we noted that the structure of the shocked zone, the region between the shock front and the contact discontinuity, can be quite complex\footnote{For a comprehensive description of the shocked jet zone see \cite{Kohler}.}. In principle, $p_{\rm s}(z)<p_{\rm e}(z)$, so that the transverse pressure gradient can focus the post-shock flow. However, since the results on the dissipation efficiency presented in \citetalias{2009MNRAS.392.1205N} are not very sensitive to the treatment of the shocked zone, we use the simple 'Model 1' from \citetalias{2009MNRAS.392.1205N} and assume that $p_{\rm s}(z)=p_{\rm e}(z)$. The main improvement is that we take a self-consistent equation of state $p=(\gamma-1)e$ with approximate adiabatic index
\be
\gamma=\frac{12p+5\rho c^2}{9p+3\rho c^2}\,,
\ee
based on \cite{2006ApJS..166..410R}.

The local dissipation efficiency is defined as
\be
\epsilon_{\rm diss}\equiv\frac{f_{\rm diss}}{f_{\rm kin,j}}\equiv\frac{f_{\rm kin,j}-f_{\rm kin,s}}{f_{\rm kin,j}}\,,
\ee
where $f_{\rm kin}=(\Gamma-1)\Gamma\beta_\perp\rho c^3$ is the kinetic energy flux density, $f_{\rm diss}$ is the dissipated energy flux density and $\beta_\perp$ is the dimensionless velocity component perpendicular to the shock front. Under the assumption of the cold upstream plasma, it can be simplified to
\be
\epsilon_{\rm diss}=\frac{\Gamma_{\rm j}-\Gamma_{\rm s}}{\Gamma_{\rm j}-1}\,.
\ee

\section{Results}
\label{sec_res}

In \citetalias{2009MNRAS.392.1205N}, we studied the dependence of the average dissipation efficiency on $\Gamma_{\rm j}$ and $\Theta_{\rm j}$ for the case of $\eta=0$, {\ie} uniform external pressure. We found that the efficiency depends sensitively on the product $\Gamma_{\rm j}\Theta_{\rm j}$. For $\Gamma_{\rm j}\Theta_{\rm j}<1$, an approximate scaling law $\epsilon_{\rm diss}\sim 6\%(\Gamma_{\rm j}\Theta_{\rm j})^2$ can be used. For $\Gamma_{\rm j}\Theta_{\rm j}>1$, very high values can be achieved.

\begin{figure}
\includegraphics[width=\columnwidth]{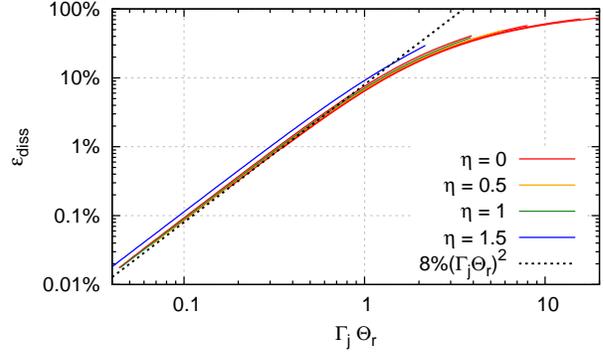}
\caption{Average dissipation efficiency as a function of the product of the jet Lorentz factor $\Gamma_{\rm j}$ and the reconfinement angle $\Theta_{\rm r}$. There are 4 families of models for the case of flat external pressure ($\eta=0$), all plotted with \emph{red solid lines}, with varying $\Theta_{\rm j}$ and $\Gamma_{\rm j}$ fixed at values 5, 10, 20 and 40. Models for $\eta>0$ are calculated using $\Gamma_{\rm j}=10$. A power-law scaling valid for $\Gamma_{\rm j}\Theta_{\rm r}<1$ is shown with the \emph{dashed black line}.}
\label{fig_eff-thetaj}
\end{figure}

Here, we have additionally calculated the average dissipation efficiency for different values of the pressure index $\eta$. The results are shown in Figure \ref{fig_eff-thetaj}. Instead of the opening angle $\Theta_{\rm j}$, the reconfinement angle $\Theta_{\rm r}$ is used, multiplied by $\Gamma_{\rm j}$, on the horizontal axis. As has been shown by \cite{1997MNRAS.288..833K}, the relation between these two angles is $\Theta_{\rm r}\sim \delta\Theta_{\rm j}$, where $\delta=1-\eta/2$. For $\eta=0$ we have $\Theta_{\rm r}\sim \Theta_{\rm j}$, so these results are consistent with the findings of \citetalias{2009MNRAS.392.1205N}. We thus generalise the previous result and show that the dissipation efficiency is determined by a single parameter in the three-dimensional space $(\Gamma_{\rm j},\Theta_{\rm j},\eta)$. In Figure \ref{fig_eff-thetaj}, we also plot a slightly different scaling law\footnote{The reason for the change of the normalising factor in the power-law scaling is that in \citetalias{2009MNRAS.392.1205N} the value for $\Gamma_{\rm j}\Theta_{\rm j}=1$ has been used, while here we require a good overall match for $0.1\lesssim\Gamma_{\rm j}\Theta_{\rm r}\lesssim 1$.}, $\epsilon_{\rm diss}=8\%(\Gamma_{\rm j}\Theta_{\rm r})^2$. A noticeable discrepancy for $\Gamma_{\rm j}\Theta_{\rm r}<1$ can only be seen for the case of $\eta=1.5$.

\begin{figure}
\includegraphics[width=\columnwidth]{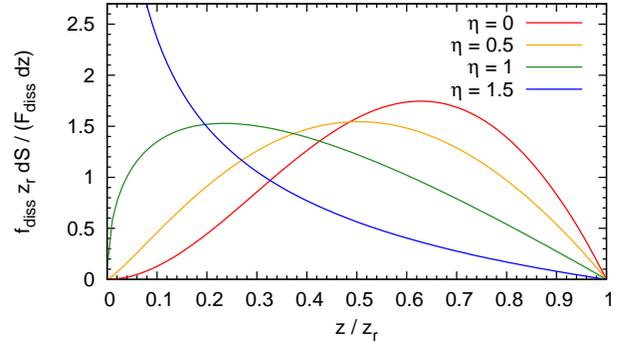}
\caption{Energy dissipation rate per unit of jet length $dz$ for different external pressure indices $\eta$. The length scale $z$ is normalised to the reconfinement length $z_{\rm r}$. The dissipation rate profiles are normalised to unity, with $F_{\rm diss}=\int f_{\rm diss}{\rm d}S$, where ${\rm d}S=2\pi r_{\rm s}{\rm d}z/\cos\alpha_{\rm s}$ is the shock front surface area.}
\label{fig_eff_dissen-z}
\end{figure}

In addition to the average dissipation efficiency, we are interested in the longitudinal distribution of the dissipated energy. Figure \ref{fig_eff_dissen-z} shows the energy flux dissipated per unit jet length $dz$ for different external pressure indices $\eta$. The profiles of dissipated energy depend very strongly on $\eta$. For the flat external pressure distribution ($\eta=0$), most of the energy is dissipated beyond the half of the reconfinement length. The amount of dissipated energy tends to 0 as $z\to z_{\rm r}$, even though the dissipation efficiency increases with $z$, since the shock becomes less and less oblique and consequently $\Gamma_{\rm s}$ decreases. But this increase in efficiency is much slower than a decrease in the jet cross-section and hence a decrease in the jet kinetic energy flux per unit $dz$. The main dissipation region shifts closer to the jet origin with increasing $\eta$. The peak of energy dissipation rate $z_{\rm diss,max}$ is located at $\sim 0.63 z_{\rm r}$ for $\eta=0$, at $\sim 0.5 z_{\rm r}$ for $\eta=0.5$ and at $\sim 0.23 z_{\rm r}$ for $\eta=1$. For $\eta=1.5$, the dissipation profile changes to monotonically decreasing with $z$ and the bulk of the dissipation takes place very close to the jet origin.

\begin{figure}
\includegraphics[width=\columnwidth]{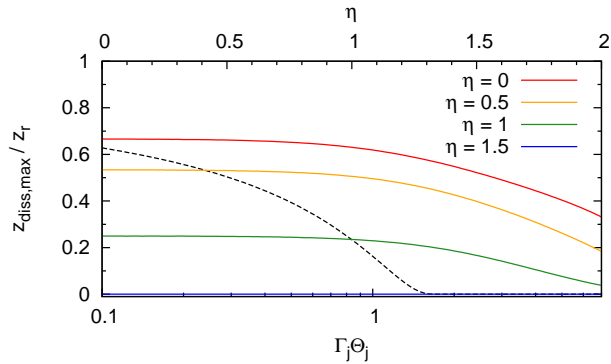}
\caption{Position $z_{\rm diss,max}$ of the peak of the energy dissipation rate along the jet axis, relative to the reconfinement length $z_{\rm r}$, as a function of the external pressure index $\eta$ and the jet opening angle $\Theta_{\rm j}$. \emph{Solid colour lines} show the dependence of $z_{\rm diss,max}/z_{\rm r}$ on the product of $\Theta_{\rm j}$ and the jet Lorentz factor $\Gamma_{\rm j}$ (plotted against the lower $x$-axis) for several values of $\eta$. \emph{Dashed black line} shows the dependence of $z_{\rm diss,max}/z_{\rm r}$ on $\eta$ (plotted against the upper $x$-axis) for $\Gamma_{\rm j}=10$ and $\Theta_{\rm j}=5^\circ$.}
\label{fig_eff_zthmax-thetaj}
\end{figure}

Figure \ref{fig_eff_zthmax-thetaj} shows the dependence of $z_{\rm diss,max}$ on the jet opening angle and the external pressure index. For a given $\eta\le 1$, the position of the dissipation peak with respect to the reconfinement length $z_{\rm r}$ is relatively stable for $\Gamma_{\rm j}\Theta_{\rm j}<1$ and is systematically shifted towards the jet origin for $\Gamma_{\rm j}\Theta_{\rm j}>1$. For $\eta=1.5$, the dissipation peak is always located very close to $z=0$. The \emph{black dashed line} shows the dependence of the location of the dissipation peak on $\eta$ for fixed $\Gamma_{\rm j}$ and $\Theta_{\rm j}$. We find that the $z_{\rm diss,max}/z_{\rm r}$ ratio decreases with $\eta$, falling to the vicinity of 0 for $\eta\gtrsim 1.3$. It is remarkable that, despite the fact that for $\eta>1$ the dissipation is concentrated close to the jet origin, the average dissipation efficiency scales with the reconfinement angle and not the opening angle. And although there are large differences in the spatial distributions of the energy dissipation rate, $\epsilon_{\rm diss}$ scales in the same way for all values of $\eta$.

\section{Discussion}
\label{sec_dis}

The results of this work and \citetalias{2009MNRAS.392.1205N} show that relativistic reconfinement shocks can be very efficient means of energy dissipation. Their efficiency depends on a simple combination of fundamental parameters of the jet and its environment. For many astrophysical jets, their Lorentz factors, opening angles and total powers can be measured or significantly constrained. In such cases it is possible to test the hypothesis that energy dissipation is dominated by the reconfinement shock.

\subsection{GRBs vs AGNs}

Achromatic breaks detected by Swift in some afterglow light curves allow one to constrain GRB jet opening angles. In several cases it has been found that $\Gamma_{\rm j}\Theta_{\rm j}\gg 1$. It became a challenge for numerists studying the initial acceleration and collimation of relativistic jets to reproduce such wide jets \citep{2009MNRAS.394.1182K}. The solution was to interrupt the collimation at some point, as would be expected for a jet breaking out of its host star \citep{2010NewA...15..749T,2010MNRAS.407...17K}. Such a situation is unique for GRB jets and allows reconfinement shocks forming at larger distances to be very efficient dissipators.

A scenario for a long GRB involving a very efficient reconfinement shock has been investigated numerically by \cite{2009ApJ...700L..47L}. The initial parameters adopted by them are $\Gamma_{\rm j}=400$ and $\Theta_{\rm j}=10^\circ$, which translates to $\Gamma_{\rm j}\Theta_{\rm j}=70$. Our model predicts in such a case an efficiency of $\sim 90\%$ for $\eta=0$ and $\sim 82\%$ for $\eta=1$. Their numerical result is thus consistent with our scaling law.

Observations of AGN jets indicate that they satisfy the relation $\Gamma_{\rm j}\Theta_{\rm j}\lesssim 1$ \cite[\eg,][]{2009A&A...507L..33P}, for which we predict at most a moderate dissipation efficiency. This is consistent with the initial jet collimation not being interrupted due to a change in the environment. The value of $\epsilon_{\rm diss}\sim 8\%$, corresponding to the case of $\Gamma_{\rm j}\Theta_{\rm r}\sim 1$, is in line with estimated radiative efficiencies of the brightest blazars. As we show below, the dissipation efficiency of jets in low-luminosity AGNs, such as M87, can be much lower.

\subsection{The jet of M87}

\cite{2006MNRAS.370..981S} provided a thorough review of the properties of the M87 jet and its environment. The Lorentz factor is estimated at $\Gamma_{\rm j}\sim 6$ and the viewing angle at $\theta_{\rm obs}\gtrsim 20^\circ$. The HST-1 knot is located at deprojected distance of $z_{\rm r}\sim 180\;{\rm pc}$. The pressure distribution of the host galaxy within $z_{\rm B}\sim 230\;{\rm pc}$ has been estimated as $p_{\rm ext}(z)=p_{\rm B}(z/z_{\rm B})^{-\eta}$, where $p_{\rm B}=1.5\times 10^{-9}\;{\rm dyn\,cm^{-2}}$ and $\eta=1.2$ (hence $\delta=0.4$). The position of the knot and the distribution of external pressure can be used to calculate the jet power, employing a formula derived from the analytical model of \cite{1997MNRAS.288..833K}:
\be
\label{eq_M87_jet_power}
L_{\rm j}=\left(\frac{\pi c p_{\rm B}}{\mu\beta_{\rm j}}\right)\left(\frac{z_{\rm r}^{2\delta}z_{\rm B}^\eta}{\delta^2}\right)\sim 5\times 10^{44}\;{\rm erg\,s^{-1}}\,,
\ee
where $\mu=17/24$. This value is a bit higher than the estimate $10^{44}\;{\rm erg\,s^{-1}}$ obtained from the energetics of the radio lobes by \cite{1996ApJ...467..597B}.

This study can be complemented by an estimate of the dissipation efficiency. We already know the jet Lorentz factor and we need to calculate the reconfinement angle. This is complicated by the fact that the jet region immediately upstream the HST-1 knot is not visible on any high-resolution radio maps. Using the 20 GHz VLBA map from \cite{2007ApJ...663L..65C}, we measure the projected aspect ratio of the jet section up to HST-1: $(2r_{\rm m}/z_{\rm r})_{\rm proj}\sim 0.026$. This corresponds to the actual aspect ratio of $(r_{\rm m}/z_{\rm r})=(r_{\rm m}/z_{\rm r})_{\rm proj}\sin\theta_{\rm obs}\sim 0.0045$. Using the analytical model of \cite{1997MNRAS.288..833K}, it is possible to express the reconfinement angle as
\be
\Theta_{\rm r}\sim (1+\delta)^{1+1/\delta}\left(\frac{r_{\rm m}}{z_{\rm r}}\right)\,.
\ee
For $\eta=1.2$, we obtain the value $\Theta_{\rm r}\sim 0.83^\circ$. Since $\Gamma_{\rm j}\Theta_{\rm r}\sim 0.087\ll 1$, the dissipation efficiency can be calculated from the approximate scaling relation: $\epsilon_{\rm diss}\sim 6\times 10^{-4}$. Multiplying it by the jet power estimated in Equation \ref{eq_M87_jet_power}, we obtain the energy dissipation rate $L_{\rm diss}=\epsilon_{\rm diss}L_{\rm j}\sim 3.3\times 10^{41}\;{\rm erg\,s^{-1}}$. If all of the dissipated energy were radiated away, the observed luminosity would be $L_{\rm obs}\sim (\mathcal{D}_{\rm j}^3/\Gamma_{\rm j})L_{\rm diss}\sim 6\times 10^{41}\;{\rm erg\,s^{-1}}$. Here, $\mathcal{D}_{\rm j}=[\Gamma_{\rm j}(1-\beta_{\rm j}\cos\theta_{\rm obs})]^{-1}\sim 2.3$ is the jet Doppler factor and the relativistic boost factor $(\mathcal{D}_{\rm j}^3/\Gamma_{\rm j})$ is used in the form that is valid for a stationary emitting region, rather than a co-moving one \citep[see][]{1997ApJ...484..108S}. At the distance of $d_{\rm L}\sim 16\;{\rm Mpc}$, the observed bolometric flux would be $f_{\rm obs}=L_{\rm obs}/(4\pi d_{\rm L}^2)\sim 2.1\times 10^{-11}\;{\rm erg\,s^{-1}\,cm^{-2}}$. This value is a bit higher than the actually observed broad-band flux of M87 \citep[see Figure 4 in][]{2009ApJ...707...55A}, which is dominated by the non-thermal emission from the inner jet. Interestingly, the observed flux could be matched if the lower estimate for the jet power by \cite{1996ApJ...467..597B} is used instead of the result of Equation \ref{eq_M87_jet_power}. This indicates that the simple analytic model of relativistic reconfinement shocks overestimates the jet power, but predicts a correct dissipation efficiency.

If the radio emission from the inner jet of M87 is produced at the reconfinement shock, its spatial distribution should be related to the distribution of the dissipation rate. The radio map from \cite{2007ApJ...663L..65C} shows that emission peaks close to the galactic nucleus \citep[see also][]{2011Natur.477..185H} and decreases monotonically with the distance. As shown in Figures \ref{fig_eff_dissen-z} and \ref{fig_eff_zthmax-thetaj}, the dissipation rate behaves in a similar manner for the external pressure index $\eta\gtrsim 1.3$. The actual index inferred for M87, $\eta=1.2$, is close to this range. Also, the radio map shows an edge-brightened jet, which is a natural consequence of dissipation at reconfinement shocks \citep{2009MNRAS.395..524N}

\section{Conclusions}
\label{sec_con}

This work generalises the findings of \citet[][Paper I]{2009MNRAS.392.1205N} on the dissipation efficiency of relativistic reconfinement shocks and sets these studies in a broader astrophysical context.

We find that the average dissipation efficiency depends on the product of the jet Lorentz factor $\Gamma_{\rm j}$ and the reconfinement angle $\Theta_{\rm r}$, which is equal to the opening angle $\Theta_{\rm j}$ for a flat distribution of external pressure ($\eta=0$). For $\Gamma_{\rm j}\Theta_{\rm r}<1$, an approximate scaling law $\epsilon_{\rm diss}\sim 8\%(\Gamma_{\rm j}\Theta_{\rm r})^2$ can be used. This moderate-efficiency regime can be applied to the jets of AGNs, while the high-efficiency regime ($\Gamma_{\rm j}\Theta_{\rm r}\gg 1$) is characteristic for GRBs. The differences in radiative efficiency between these sources may be related to the different circumstances of the initial jet collimation process. A similar idea for the unification of relativistic jets between GRBs and AGNs has been recently formulated by \cite{2011Nemmen}.

Our results have been applied to the jet of radio galaxy M87, hosting a peculiar knot HST-1. Emission from the inner jet of M87 is consistent with dissipation at a reconfinement shock extending upstream from HST-1 in two aspects:
\begin{itemize}
\item[--]
the broad-band luminosity of M87 is consistent with the product of the dissipation efficiency $\epsilon_{\rm diss}$ predicted by our model and the independently estimated jet power;
\item[--]
radio emission peaking close to the nucleus resembles the dissipation profile of reconfinement shocks for external pressure index $\eta=1.2$.
\end{itemize}

Our results suggest that reconfinement shocks may be a dominant dissipation mechanism in astrophysical relativistic jets. As such, they deserve more attention and more detailed investigations.

\section*{Acknowledgements}
The author is grateful to Marek Sikora for his advise and support. Mitch Begelman and Kris Beckwith read the manuscript and provided helpful comments.
This work has been partly supported by the Polish MNiSW grants N~N203 301635 and N~N203 386337, the Polish ASTRONET grant 621/E-78/SN-0068/2007, the NSF grant AST-0907872 and the NASA Astrophysics Theory Program grant NNX09AG02G.

\end{document}